\documentstyle[aps,twocolumn]{revtex}
\begin{document}

\draft

\title{Quantum information processing using quantum dot spins and cavity-QED}

\author{A. Imamo\=glu$^{1,2}$, D.~D. Awschalom$^2$, G. Burkard$^3$,
D.~P. DiVincenzo$^3$, D. Loss$^3$, M. Sherwin$^2$, A. Small$^2$}

\address{$^1$ Department of Electrical and Computer Engineering,
University of California, Santa Barbara, CA 93106}

\address{$^2$ Department of Physics,
University of California, Santa Barbara, CA 93106}

\address{$^{3}$
Department of Physics and Astronomy,
University of Basel, Klingelbergstrasse 82,
CH-4056 Basel, Switzerland}

\address{$^{4}$
IBM Research Division, T.J.\ Watson Research Center,
P.O.\ Box 218, Yorktown Heights, NY 10598}

\date{\today}

\twocolumn[\hsize\textwidth\columnwidth\hsize\csname @twocolumnfalse\endcsname

\maketitle

\begin{abstract}
The electronic spin degrees of freedom in semiconductors typically
have decoherence times that are several orders of magnitude longer
than other relevant timescales. A solid-state quantum computer based
on localized electron spins as qubits is therefore of potential
interest. Here, a scheme that realizes controlled interactions between
two distant quantum dot spins is proposed. The effective long-range
interaction is mediated by the vacuum field of a high finesse
microcavity. By using conduction-band-hole Raman transitions induced
by classical laser fields and the cavity-mode, parallel controlled-not operations
and arbitrary single qubit rotations can be realized. Optical
techniques can also be used to measure the spin-state of each quantum
dot.
\end{abstract}



\vskip2pc]
\narrowtext

Within the last few years, quantum computation (QC) has developed into
a truly interdisciplinary field involving the contributions of physicists,
engineers, and computer scientists \cite{Steane98}.  The seminal discoveries of
Shor and others, both in developing quantum algorithms for important
problems like prime factorization \cite{Shor94}, and in developing
protocols for
quantum error correction (QEC) \cite{Shor95} and fault-tolerant quantum
computation \cite{Shor96}, have indicated the desirability and the
ultimate feasibility of the experimental realization of QC in various
quantum systems.

The elementary unit in most QC schemes is a two-state system referred
to as a quantum bit (qubit). Since  QEC can only work if the decoherence rate is
small, it is crucial to identify schemes where the qubits are well isolated from
their environment.  Ingenious schemes based on Raman-coupled low-energy states
of trapped ions \cite{Cirac95} and nuclear spins in chemical solutions
\cite{Cory97} satisfy this criterion, in addition to providing
methods of
fast quantum manipulation of qubits that do not introduce significant
decoherence. Even though these schemes are likely to provide the first examples
of quantum information processing at 5-10 qubit level, they do not appear to be
scalable to larger systems containing more than 100 qubits.

Here, we propose a new scheme for quantum information processing based on
quantum dot (QD) electron spins coupled through a microcavity mode. The
motivation for this scheme is threefold: (1) a QC scheme based on semiconductor
quantum dot arrays should be scalable to $\ge 100$ coupled qubits;
(2) recent experiments demonstrated very long spin decoherence times for
conduction band electrons in III-V and II-VI semiconductors \cite{Kikkawa97},
making electron spin a likely candidate for a qubit; and (3) cavity-QED
techniques can provide long-distance, fast interactions between qubits
\cite{Pellizzari95}. The QC scheme detailed below relies on the use of a single
cavity mode and laser fields to mediate coherent interactions between
distant QD spins. As we will show shortly, the proposed scheme does not require
that QDs be identical and can be used to carry out parallel quantum logic
operations \cite{Molmer99}.

We note that a QC scheme based on electron spins in QDs have been
previously proposed \cite{Loss98}: this scheme
is based on local exchange interactions controlled by 
electrodes. The possibility of coherently manipulating motional
degrees of freedom of QD electrons using terahertz cavity-QED has also
been discussed \cite{Sherwin99}. Other all-solid-state QC schemes
include the scheme by Kane to implement NMR in doped silicon wafers
\cite{Kane98}. The principal feature that distinguishes the present
proposal from its predecessors is the use of all-optical Raman
transitions to couple two conduction-band spin states: this
potentially enables us to combine the ultra-long spin coherence times
with fast, long-distance, parallel optical switching.

The proposed scheme is detailed in Figure 1: the doped QDs are
embedded in a microdisk structure with diameter $d \simeq 2 \mu$ and
thickness $d \simeq 0.1
\mu$. Motivation for choosing this structure over other alternatives comes
from recent experiments demonstrating that InAs self-assembled QDs can be
embedded in microdisk structures with a cavity quality factor $Q \simeq
12,000$
\cite{Gerard99}.  We assume that the QDs are designed such that the quantum
confinement along the z-direction is the strongest: this is the case
both in QDs defined by electrical gates and in self-assembled QDs. The
in-plane confinement is also assumed to be large enough to guarantee
that the electron will always be in the ground-state orbital.  Because
of the strong z-azis confinement, the lowest energy eigenstates of
such a III-V or II-VI semiconductor QD consist of $|m_z=\pm1/2\rangle$
conduction-band states and $|m_z=\pm3/2\rangle$ valence-band
states. The QDs are doped such that each QD has a full valence band
and a single conduction band electron: we assume that a uniform
magnetic field along the x-direction ($B_x$) is applied, so the QD
qubit is defined by the conduction-band states $|m_x=-1/2\rangle =
|\!\!\downarrow\rangle$ and $|m_x=1/2\rangle = |\!\!\uparrow\rangle$
(Fig.~2). The corresponding energies are
$\hbar \omega_{\downarrow}$ and $\hbar \omega_{\uparrow}$, respectively. 

One of the key elements of the proposed scheme is the Raman coupling
of the two spin eigenstates via strong laser fields and a single-mode
microcavity mode.  Since the QD array dimensions are assumed to be on
the order of several microns, we assume that the laser fields are
coupled selectively to given QDs using near-field techniques; e.g. via
tapered fiber tips. The 1-bit operations proceed by applying two laser
fields $E_{L,x}(t)$ and $E_{L,y}(t)$ with Rabi frequencies 
$\Omega_{L,x}$ and $\Omega_{L,y}$, and frequencies $\omega_{L,x}$
and $\omega_{L,y}$ (polarized along the x and y directions,
respectively) that exactly satisfy the Raman-resonance condition
between $|\!\!\downarrow\rangle$ and $|\!\!\uparrow\rangle$.  The
laser fields are turned on for a short time duration that satisfies a
$\pi/r$-pulse condition, where $r$ is any real number. The process can
be best understood as a Raman $\pi/r$-pulse for the {\sl hole} in the
conduction band state. The laser field polarizations
should have non-parallel components in order to create a non-zero
Raman coupling (if there is no heavy-hole light-hole mixing). These
arbitrary single-bit rotations can naturally be carried out in parallel. In
addition, the QDs that are not doped by a single electron never couple to the Raman
fields and can safely be ignored.

The fundamental step in the implementation of 2-qubit operations is the
selective coupling between the {\sl control} and {\sl target} qubits that is
mediated by the microcavity field. To this end, we assume that the cavity-mode
with energy $\hbar \omega_{cav}$ (assumed to be x-polarized) and a laser field
(assumed to be y-polarized) establishes the Raman transition between the two
conduction-band states, in close analogy with the atomic cavity-QED schemes
\cite{Pellizzari95}.  After eliminating the valence band
states \cite{Madelung}, we obtain the effective Hamiltonian ($\hbar=1$)
\begin{eqnarray}
\hat{H}_{eff} & = & \omega_{cav} \hat{a}_{cav}^{\dagger}
\hat{a}_{cav} + \sum_i \left[\omega_{\uparrow \downarrow}^i
\hat{\sigma}_{\uparrow \uparrow}^i
- \frac{g_{cav}^2}{ \Delta \omega_{\uparrow}^i}
 \, \hat{\sigma}_{\downarrow \downarrow}^i \,  \hat{a}_{cav}^{\dagger}
\hat{a}_{cav} \right. \nonumber \\
& - &  \left. \frac{(\Omega_{L,y}^i)^2}{\Delta \omega_{\downarrow}^i}
\hat{\sigma}_{\uparrow \uparrow}^i \, + \, i  g_{eff}^i \left[
\hat{a}_{cav}^{\dagger}
\hat{\sigma}_{\downarrow \uparrow}^i e^{-i \omega_{L,y}^i t} - h.c.\right]
\right],
\label{eq1}
\end{eqnarray} where
\begin{eqnarray} g_{eff}^i(t) & = & \frac{g_{cav}
\Omega_{L,y}^i(t)}{2} \left(\frac{1}{\Delta \omega_{\uparrow}^i}  +
\frac{1}{\Delta \omega_{\downarrow}^i}\right)
\label{eq2}
\end{eqnarray}
is the effective 2-photon coupling coefficient for the spins of the
${\rm i^{th}}$ QD. $\hat{a}_{cav}$ denotes the cavity mode
annihilation operator and $\hat{\sigma}_{\uparrow \downarrow}^i =
|\!\!\uparrow\rangle \langle \downarrow \!\!|$ is the spin projection
operator for the
${\rm i}^{th}$ QD. The exact two-photon-resonance condition would be $\Delta
\omega_{\uparrow}^i =\omega_{\uparrow}^i - \omega_v^i - \omega_{cav} = \Delta
\omega_{\downarrow}^i  = \omega_{\downarrow}^i - \omega_v^i - \omega_L^i$.
$\hbar \omega_{v}^i$ denotes the energy of the relevant valence band
states\cite{foot1}, and $\omega_{\uparrow \downarrow}^i =
\omega_{\uparrow}^i -\omega_{\downarrow}^i$. The derivation of
$\hat{H}_{eff}$ assumes $\Delta\omega_{\uparrow,\downarrow}^i \gg
g_{cav}$, $\omega_{\uparrow \downarrow}^i \gg k_B T$,
and $\omega_{\uparrow\downarrow}^{i,j} \gg g_{eff}^i > \Gamma_{cav}$,
where $\Gamma_{cav}$ denotes the cavity decay rate (not included in
Eq.~(\ref{eq1})). The third and fourth terms of
Eq.~(\ref{eq1}) describe the ac-Stark-effect caused by the cavity and
laser fields, respectively.

The first step in the implementation of a CNOT operation would be to
turn on laser fields $\omega_L^i$ and $\omega_L^j$ to establish near
two-photon resonance condition for both the control (i) and the target
(j) qubits:
\begin{eqnarray}
\Delta_i & = & \omega_{\uparrow \downarrow}^i -
\omega_{cav} + \omega_L^i \; =  \; \Delta_j  \ll
\omega_{\uparrow \downarrow}^{i,j} \;\;\;.
\label{eq3}
\end{eqnarray}
If we choose two-photon detunings $\Delta_i$ large compared to the
cavity linewidth and $g_{eff}^i(t)$,
we can eliminate the cavity-degrees-of-freedom \cite{Madelung} to
obtain an effective two-qubit interaction Hamiltonian in the rotating
frame (interaction picture with $H_0 = \sum_i
\omega^i_{\uparrow\downarrow}\hat{\sigma}^i_{\uparrow\uparrow}$):
\begin{eqnarray}
\hat{H}_{int}^{(2)} & = & \sum_{i \neq j} \tilde{g}_{ij}(t)
\left[ \hat{\sigma}_{\uparrow
\downarrow}^i \hat{\sigma}_{\downarrow \uparrow}^j e^{i\Delta_{ij}t} \, + \,
\hat{\sigma}_{\uparrow
\downarrow}^j \hat{\sigma}_{\downarrow \uparrow}^i e^{-i\Delta_{ij}t} \right] \;\;\;,
\label{eq4}
\end{eqnarray}
where $\tilde{g}_{ij}(t) = g_{eff}^i(t) g_{eff}^j(t)/ \Delta_i$ and $\Delta_{ij} =
\Delta_i - \Delta_j$. Equation~(\ref{eq4}) is
one of the principal results of this Letter: it shows that Raman
coupling via a common cavity mode can establish fully controllable
long-range transverse spin-spin interactions between  two distant QD electrons, by
choosing $\omega_L^i$ and $\omega_L^j$ such that $\Delta_{ij} = 0$. 
We can see from $\hat{H}_{int}^{(2)}$ that the size non-uniformity of QDs is in
principle completely irrelevant for the proposed scheme: small differences in
g-factor between different QDs can be adjusted by the choice of $\omega_L^i$ and
$\omega_L^j$. The additional single-spin terms of the effective interaction
Hamiltonian (not shown in Eq.~(\ref{eq4})) induce single
qubit phase-shifts; however, these terms can be safely discarded since
they are smaller than the ac-Stark terms of Eq.~(\ref{eq1}).

Next, we turn to the implementation of the conditional phase-flip (CPF)
operation between spins $i$ and $j$: to
this end, we set $\Delta_{ij}=0$ and obtain
\begin{eqnarray}
\hat{H}_{int,ij}^{(2)} & = &  \frac{\tilde{g}_{ij}(t)}{2}
\left[ \hat{\sigma}_{y}^i \hat{\sigma}_{y}^j \, + \,
\hat{\sigma}_{z}^i \hat{\sigma}_{z}^j \right] \;\;\;,
\label{eq5}
\end{eqnarray}
where $\hat{\sigma}_{y}$ and $\hat{\sigma}_{z}$ are the Pauli operators. In
this form, we recognize that the effective interaction Hamiltonian between two
QDs is that of transverse spin-spin coupling. Previously, the possibility of 
carrying out universal QC has been shown for Hamiltonians  of the form $H \sim J
{\bf S}^i \cdot {\bf S}^j$ \cite{Loss98}. For the transverse spin-spin coupling of
Eq.~(\ref{eq5}), we find that a non-trivial two-qubit gate, such as the
conditional-phase-flip (CPF) operation, can be carried out by
combining $\hat{H}_{int,ij}^{(2)}$ with one-bit rotations.
The unitary evolution operator under the Hamiltonian of Eq.~(\ref{eq5}) is
\begin{eqnarray}
\hat{U}_{XY}(\phi)
& = &  T \exp\left[ i \int dt \hat{H}_{int,ij}^{(2)}\right]
\;\;\;,
\label{eq6}
\end{eqnarray}
where $\phi = \int dt \tilde{g}_{ij}(t)$. The CPF gate ($\hat{U}_{CPF}$) can be
realized by the sequence of operators
\begin{eqnarray}
\hat{U}_{CPF} & = &  e^{i \pi/4} e^{i \pi {\bf n}_i \cdot
{\mbox{\boldmath $\sigma$}}_i/3} e^{i \pi {\bf n}_j \cdot
{\mbox{\boldmath $\sigma$}}_j/3}
\hat{U}_{XY}(\pi/4) \nonumber \\
& &\times  e^{i \pi \sigma_z^i/2} \hat{U}_{XY}(\pi/4)
e^{i \pi \sigma_y^i/4} e^{i \pi \sigma_y^j/4}
\;\;\;,
\label{eq7}
\end{eqnarray}
where ${\mbox{\boldmath $\sigma$}}$ denotes the vector Pauli operator,
${\bf n}_i=(1,1,-1)/\sqrt{3}$, and ${\bf n}_j=(-1,1,1)/\sqrt{3}$.
Note that all operators in this sequence are understood in the
interaction picture defined above in Eq.~(\ref{eq4}). The controlled-not
gate can be realized by combining the CPF operation with single-qubit
rotations $\hat{U}_{CNOT} = \exp[ -i \pi \sigma_z^j/4] \,  \hat{U}_{CPF} \,
\exp[ i \pi \sigma_z^j/4]$.

Equation~(\ref{eq4}) also indicates that two-qubit interactions such as the CPF
operation can be carried out in parallel. To see how this works, we consider for
simplicity 4 QDs where we set $\Delta_a = \Delta_c$ and $\Delta_b = \Delta_d$ and
choose
$\Delta_{ab} \gg \tilde{g}_{ij}(t)$ ($i,j = a,b,c,d$;  $i \neq j$), by adjusting the
corresponding laser frequencies (Fig.~1). For these parameters, QDs a and c (as well
as b and d) will couple to each other via the transverse spin-spin interaction of
Eq.~(\ref{eq5}), whereas the coupling between all other pairings of QDs will be
energy non-conserving and average out to zero. Generalizing this procedure, we find
that a single cavity-mode can be used to carry out many parallel 2-qubit operations.
An analogous method to achieve parallel operations in ion-trap QC was  described
in Ref.~\cite{Molmer99}.

The time required to complete a 2-qubit gate operation in this scheme will
be limited by the strength of the electron-hole-cavity coupling.  1-bit
operations can be completed in $10\,{\rm psec}$ time scale, assuming
$\Omega_{L,x},\Omega_{L,y} \simeq 1\,{\rm meV}$, and
$\Delta \omega_{\uparrow}^i \simeq \Delta \omega_{\downarrow}^i \simeq 5\,{\rm
meV}$. If we assume $g_{cav} \simeq
\Delta_i \simeq 0.5\,{\rm meV}$, we find $\tilde{g}_{ij} \simeq 0.02
\,{\rm meV}$ and that CPF operation can be carried out in $\sim 100 \,{\rm psec}$.
For InAs self-assembling QDs,
$g_{cav} \simeq 0.1 \,{\rm meV}$ for a typical cavity volume $V_{cav} = 4 (2
\pi c_n /
\omega_{cav})^3$, where $c_n$ is the speed of light in the medium.
It should be possible to obtain $g_{cav} \simeq 0.5\,{\rm meV}$ by  utilizing
large area QDs \cite{foot2}. 

One of the key issues in the feasability of a quantum information processing
scheme is the relative magnitude of the decoherence rates as compared to the
gate-operation time. As indicated earlier, our scheme is motivated by the
 $1 \, \mu sec$ long coherence times of conduction band electrons observed
in doped QW and bulk semiconductors \cite{Kikkawa97}. Recent experiments in
undoped QDs on the other hand indicate that spin decoherence times are at least
as long as 3 nsec even in the presence of valence band holes \cite{Gupta99}; 
the corresponding times for undoped QW structures is on the
order of $50 \, {\rm psec}$. It should therefore be safe to assume
that at least in an ideal system such as GaAs QDs embedded in AlGaAs,
the spin decoherence times will be around $1 \, \mu {\rm sec}$. 

The principal spin decoherence mechanism in QDs is the strong
spin-orbit interaction in the valence-band. Presence of (valence-band) holes
therefore could potentially increase the decoherence rate by several orders of
magnitude. By utilizing the Raman scheme, we only create {\sl virtual holes}
with a probability of $\simeq 0.01$ for the assumed values of $\Omega_{L,y},
\Delta \omega_{\uparrow}^i, \Delta \omega_{\downarrow}^i$. If the hole-spin
decoherence time is $1 \, {\rm nsec}$, this could give rise to an effective
decoherence time of $100 \, {\rm nsec}$; we reiterate however that this
gate error only appears in the presence of the laser fields. In the
cavity-QED scheme that we propose, the finite cavity lifetime ($
\Gamma_{cav}^{-1} \sim 10 \, {\rm psec}$) also introduces a decoherence
mechanism.
However, since the cavity-mode is also only virtually excited during the
transverse spin-spin interaction (with probability $\sim 0.01$ for the assumed
parameters), the effective decoherence time will be on the order of $1 \,
{\rm nsec}$.
Therefore, the primary technological limitation for the proposed scheme is the
relatively fast photon loss rate. This rate can be improved either by
new processing techniques for microdisk structures, or by coupling QDs to 
ultra-high finesse cavities of silica microspheres \cite{Wang99}.

Finally, the accurate measurement of the spin state of each qubit is an
essential requirement in a QC scheme. In our system, this can be achieved by
applying a laser field $E_{L,y}$ to the QD to be measured, in order to realize
exact two-photon resonance with the cavity mode. If the QD spin is in state
$|\!\!\downarrow\rangle$, there is no Raman coupling and no photons will be
detected. If on the other hand, the spin state is $|\!\!\uparrow\rangle$, the
electron will exchange energy with the cavity mode and eventually a single
photon will leak out of the cavity. A single photon
detection capability is sufficient for detecting a single spin. 

One of the key assumptions of our scheme is the need to address each QD
selectively. Using state-of-the-art near-field techniques can give us
resolution that is on the order of $1000 \, {\rm \AA}$, which in turn gives an
upper limit of
$\sim 20$ as the total number of QDs that can be coupled to a single
cavity-mode. An alternative method to couple larger number of QDs could be to
use the electric field dependence of the electron g-factor $g_{elec}$ in
semiconductors \cite{g-factor}. An ideal situation for the proposed scheme is to
design the system such that each QD will have a dedicated gate electrode that can be
used to switch its $g_{elec}$ and hence bring the QD in and out of
Raman resonance with the laser and cavity-modes \cite{Sherwin99}. The laser fields
can still be pulsed, but need not be localized spatially. 

In summary, we have described a new quantum information processing scheme based
on QDs strongly coupled to a microcavity mode. The primary achievement of this
scheme is the realization of parallel, long-range transverse spin-spin interactions
between conduction band electrons, mediated by the cavity mode. Such interactions
can be used to carry out quantum computation. Currently, the primary
technological limitation of the proposed scheme is the short photon lifetime in
state-of-the-art microcavities. Further improvements on the ratio of gate
operation time to decoherence times may be achieved using adiabatic passage
schemes to eliminate or minimize the virtual amplitude generated in the
cavity-mode and the valence-band hole states. The Hamiltonian of Eq.~(\ref{eq4})
can also be used to achieve quantum state transfer from one QD to another, along the
lines described in the context of atomic cavity-QED.

We thank J.-M.~ Gerard and B.~Gayral for sharing their results on
microdisks before publication. A.~I. would also like to thank A. Sorensen and K.
Molmer for bringing to his attention Ref.~\cite{Molmer99}, which was very helpful
in demonstrating parallel two-qubit operations. This work is supported by the
grants ARO DAAG55-98-1-0366 and DAAG55-98-C-0041. A.~I. acknowledges the support
of a David and Lucile Packard Fellowship. G.~B. and D.~L. acknowledge
funding from the Swiss National Science Foundation. A. S. acknowledges a NSF
graduate fellowship.

\begin{figure} \caption{The quantum information processing scheme that is based
on quantum dots embedded inside a microdisk structure. Each
quantum dot is addressed selectively by a laser field from a fiber-tip. The laser
frequencies are chosen to select out the pair of quantum dots that will participate
in gate operation. All dots strongly couple to a single cavity-mode.}
\label{fig1}
\end{figure}

\begin{figure} \caption{The relevant energy levels of a III-V (or II-VI)
semiconductor quantum dot. It is assumed that confinement along the z-direction
is strongest.}
\label{fig2}
\end{figure}


\begin{references}
\bibitem{foot1} Due to the quantum confinement induced splitting
between the $|m_z=\pm1/2\rangle$ and $|m_z=\pm3/2\rangle$ states, the
applied field can only slightly shift the energies of the
$|m_z=\pm3/2\rangle$ states: we neglect this shift and use a single value
$\omega_v^i$. This assumption is justified provided that the energy shift is
much smaller than $\Delta \omega_{\uparrow}^i$ and $\Delta
\omega_{\downarrow}^i$.
\bibitem{foot2} We point out that strong
coupling of a QD to a single cavity mode, which is a principal feature of the
proposed scheme, is still not observed experimentally. However, recent experiments 
in QDs coupled to microdisks \cite{Gerard99} and silica microspheres \cite{Wang99}
indicate that strong coupling could be observed in the near future.
\end{references}
\end{document}